\documentclass[prx, letterpaper, superscriptaddress, amsmath, amssymb, amssymb, reprint, floatfix]{revtex4-2}
\usepackage{graphicx}
\usepackage{dcolumn}
\usepackage{soul}
\usepackage{longtable}
\usepackage{layouts}

\usepackage{bm}
\usepackage{xcolor} 
\usepackage[normalem]{ulem}

\begin{document}

\title{RF-Squad: A radiofrequency simulator for quantum dot arrays }

\author{Tara Murphy}
\email{tm763@cam.ac.uk}
 \affiliation{Quantum Motion, 9 Sterling Way, London, N7 9HJ, United Kingdom}
 \affiliation{Cavendish Laboratory, University of Cambridge, J.J. Thomson Avenue, CB3 0HE, United Kingdom}
\affiliation{Department of Applied Mathematics (DAMPT), University of Cambridge, J.J. Thomson Avenue, CB3 0HE, United Kingdom}

\author{Katarina Brlec}
 \affiliation{Quantum Motion, 9 Sterling Way, London, N7 9HJ, United Kingdom}

\author{Giovanni Oakes}
 \affiliation{Quantum Motion, 9 Sterling Way, London, N7 9HJ, United Kingdom}

\author{Lorenzo Peri}
 \affiliation{Quantum Motion, 9 Sterling Way, London, N7 9HJ, United Kingdom}
  \affiliation{Cavendish Laboratory, University of Cambridge, J.J. Thomson Avenue, CB3 0HE, United Kingdom}

\author{Henning Sirringhaus}
 \affiliation{Cavendish Laboratory, University of Cambridge, J.J. Thomson Avenue, CB3 0HE, United Kingdom}

\author{Henry Moss}
 \affiliation{Department of Applied Mathematics (DAMPT), University of Cambridge, J.J. Thomson Avenue, CB3 0HE, United Kingdom}
  \affiliation{School of Mathematical Sciences, Lancaster University, LA1 4YW, United Kingdom}

\author{M. Fernando Gonzalez Zalba}
 \email{fernando@quantummotion.tech}
 \affiliation{Quantum Motion, 9 Sterling Way, London, N7 9HJ, United Kingdom}
 \affiliation{CIC nanoGUNE Consolider, Tolosa Hiribidea 76, E-20018 Donostia-San Sebastian, Spain}
\affiliation{IKERBASQUE, Basque Foundation for Science, E-48011 Bilbao, Spain}

  \author{David Wise}
 \affiliation{Quantum Motion, 9 Sterling Way, London, N7 9HJ, United Kingdom}

\date{\today}

\begin{abstract}

Spins in semiconductor quantum dots offer a scalable approach to quantum computing; however, precise control and efficient readout of large quantum dot arrays remain challenging, mainly due to the hyperdimensional voltage space required for tuning multiple gates per dot. To automate this process, large datasets are required for testing and training autotuning algorithms. To address the demand for such large datasets, we introduce RF-Squad, a physics-based simulator designed to realistically replicate radiofrequency (RF) reflectometry measurements of quantum dot arrays, with the ability to go beyond the Constant Interaction Model (CIM) and simulate physical phenomena such as tunnel coupling, tunnel rates, and quantum confinement. Implemented in JAX, an accelerated linear algebra library, RF-Squad achieves high computational speed, enabling the simulation of a 100$\times$100 pixel charge stability diagram of a double quantum dot (DQD) in 52.1 $\pm$0.2 milliseconds at the CIM level. Using optimization algorithms, combined with it's layered architecture, RF-Squad allows users to balance physical accuracy with computational speed, scaling from simple to highly detailed models.

\end{abstract}

\maketitle


\section{Introduction}  
\label{sec:Introduction}

Semiconductor quantum dots (QDs) have emerged as a promising platform for scalable quantum computing, demonstrating high operating fidelities and compatibility with existing semiconductor fabrication techniques \cite{gonzalezzalba2021,Arquer2021,Scappucci2021,Sreejith2024, Thomas2025,  Guidoreview}, while offering a path toward large-scale integration. Small QD arrays have been explored and measured in laboratory settings, showing promising results in their potential to act as qubits \cite{Philips2022, Mills2022, Zwerver2022, Noiri2022, Weinstein2023, Thorvaldson2025}. However, as QD arrays grow in size, the complexity of tuning, control, and readout increases dramatically, making manual tuning infeasible. Extracting meaningful information from measurements requires robust, automated approaches. Recent work has demonstrated machine learning and autotuning algorithms capable of automating QD tuning \cite{Ares2016, Schuff_2023, Zwolak2020, Moon2020, Ziegler2023}, but these methods have mainly been applied to smaller systems or specific device architectures. To develop robust generalizable algorithms for large-scale QD arrays, realistic QD array simulation environments (a digital twin), as well as large datasets of measurements, are needed  \cite{zwolak2018qflow, Zwolak2024}. Obtaining such datasets experimentally is impractical due to the sheer volume of measurements, and a fast, physics-based simulator capable of generating realistic QD measurements efficiently is required. 

In this paper, we introduce RF-Squad, a simulator designed to fill this gap by providing a method to create synthetic charge measurements of QD arrays at high speed. RF-Squad is not the first effort in developing a QD simulator. Recent advancements of simulators include QDSim \cite{qdsim} and QDarts \cite{qdarts}, both of which are optimized to rapidly produce charge maps of large QD arrays, and Qarray \cite{qarray}, which uses sophisticated algorithms to efficiently sample the Fock space for energy configuration calculations. Other examples include Pu. et. al. \cite{Pu2023}, who have taken a more targeted approach, focusing on smaller QD arrays. Although these are computationally slower, they excel in capturing a broader range of physical phenomena, providing deeper insight into experimental features. Existing simulators generally fall into one of two categories: those that provide highly realistic simulations but are computationally expensive, and those that prioritize speed at the expense of physical realism. 

RF-Squad balances both aspects in two ways: (i) it uses JAX \cite{jax2018github}---an accelerated linear algebra library---for efficient computations and employs optimization techniques to maximize efficiency. Second, it has a layered architecture, allowing users to tailor simulations based on their specific needs, choosing between computational speed and physical complexity. In doing so, it incorporates key quantum mechanical effects, such as tunnel coupling, quantum confinement effects, and voltage-dependent parameters to ensure realistic results.  Moreover, (ii) it focuses on radiofrequency (RF) measurements, the most widely used method for fast high-sensitivity readout of QD systems, providing a scalable alternative to conventional dc charge-sensing methods \cite{Gonzalez-Zalba2015, West2018, vigneau2022, ciriano2021spin, Ibberson2021, urdampilleta2019gate, zheng2019rapid, Oakes2023}. To the best of our knowledge, no other simulator provides the same level of accuracy in capturing the nuances of RF reflectometry introduced by the finite probing frequency.

In this paper, we explain each layer of complexity of RF-Squad, benchmark its performance and demonstrate its ability to replicate experimental data while maintaining computational efficiency. We aim to establish RF-Squad as the next-generation QD simulator, providing a tool that not only supports experimental research but also accelerates the development of large-scale datasets for machine learning tuning applications and use as a digital twin. 

The paper is structured as follows: In Section \ref{sec:CIM}, we explain the Constant Interaction Model, the foundational theoretical framework of RF-Squad. In Section \ref{sec:Simulation}, we break down the layered structure of RF-Squad, explaining how each physics-based layer contributes to simulating realistic QD measurements. Section \ref{sec:reduced_fock_states} explains our use of optimisation algorithms to reduce the time required for simulations while maintaining accuracy. Section \ref{sec:benchmarking} evaluates the computational efficiency of RF-Squad, analyzing the impact of different layers on the simulation speed. Finally, Section \ref{sec:examples} showcases practical applications, presenting various simulation examples and comparing their results with experimental data.

\section{The Constant Interaction Model}
\label{sec:CIM}

The Constant Interaction Model (CIM) simplifies the description of QD systems by representing them as a network of capacitance elements between QDs and applied gate voltages. These capacitive interactions are structured into two key matrices:

\begin{itemize}
    \item $\mathbf{C_{dd}}$ – the capacitance matrix describes the QD total capacitances and the coupling between QDs.
    \item $\mathbf{C_{dg}}$ – the capacitance matrix governs interactions between QDs and applied gate voltages.
\end{itemize}

\begin{figure}
    \centering
    \includegraphics[width=0.6\linewidth]{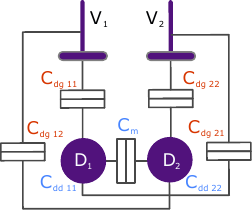}
    \caption{Schematic of a double quantum dot (DQD) in the CIM framework. The two QDs, labelled $D_1$ and $D_2$, are controlled by gate voltages $V_1$ and $V_2$. The relevant capacitance elements forming the CIM capacitance matrices are indicated: red terms represent dot-to-gate capacitance, while blue terms denote dot-to-dot capacitance. The mutual capacitance between the QDs is denoted as $C_m$ and corresponds to the elements $C_\text{{dg}, 12 (21)}$.}
    \label{fig:CIM_schematic}
\end{figure}

This framework provides a straightforward means of determining the energy state and corresponding charge configuration of a QD array. A schematic representation of this model for a DQD is shown in Fig. \ref{fig:CIM_schematic}.

The primary use case of the CIM is its ability to describe charge stability diagrams (CSDs), which are essential tools for understanding charge configurations in QD devices. In a typical CSD, gate voltages are first set to a specific value and applied to the QD array. Typically, two gate voltages are swept while others remain fixed, revealing distinct regions in voltage space that correspond to different charge occupancies of the QDs.

To quantitatively determine these charge occupancies, we compute the energy of the system based on its charge configuration. The electrochemical potential $E$ of the QD system is defined as \cite{Wiel2002}:

\begin{equation}
    E = \frac{1}{2} \vec{V}^{T} \mathbf{C^{-1}_{dd}} \vec{V},
    \label{eqn:energy_cim}
\end{equation}

\noindent where we define $\vec{V} = e (\mathbf{C_{dg}} \vec{V_g} - \vec{N})$, where $e$ is the elementary charge, $\vec{N}$ signifies the number of charges on each QD, known as the Fock state, and $\vec{V_g}$ denotes the applied gate voltages.

\begin{figure*}
    \centering
    \includegraphics[width=\textwidth]{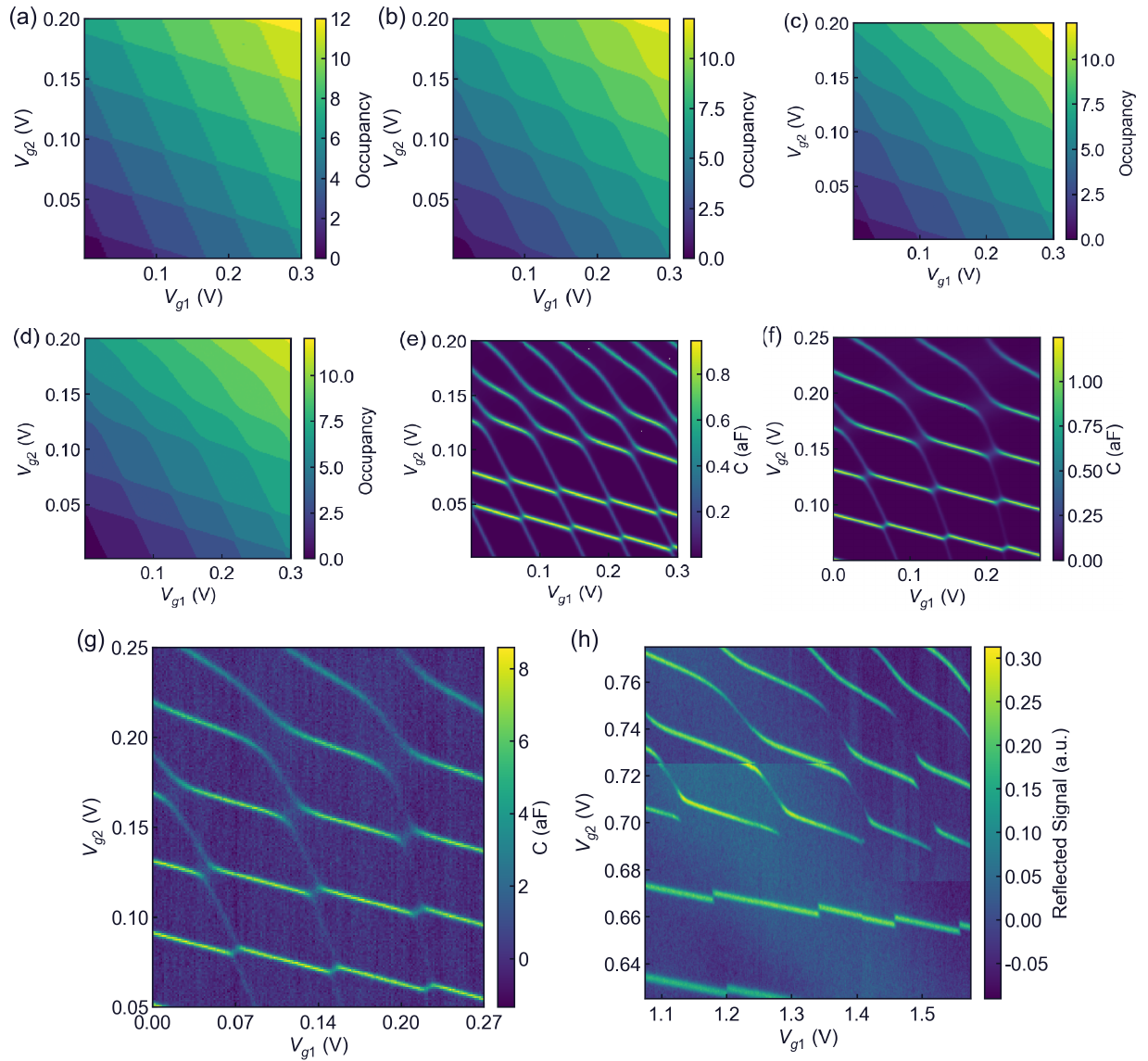} 
    \caption{Illustration of the layered architecture of RF-Squad, demonstrating how each layer progressively incorporates additional physical effects to better replicate the experimental charge stability diagram (CSD) of a DQD, shown in (h). (a) The base layer is modelled using the Constant Interaction Model (CIM). (b) Incorporates tunnel coupling by solving the eigenvalue problem in Eq~\ref{eqn:hamiltonian}, introducing curvature at charge transitions. (c) Includes quantum confinement effects, leading to periodic variations in honeycomb sizes due to additional energy contributions as described in Eq~\ref{eqn:fd_energy}. (d) Applies the Wentzel-Kramers-Brillouin (WKB) approximation to tunnel coupling, capturing the voltage-driven transition from a double to a single QD. (e) Models capacitance matrices as voltage-dependent functions, leading to non-repeating honeycomb patterns. (f) Simulates RF reflectometry using Eq~\ref{eqn:BIG}, focusing on capacitance variations induced by an oscillatory gate voltage. (g) Adds Gaussian and $1/f$ noise to replicate experimental imperfections. (h) Presents the experimentally measured CSD, serving as the benchmark for comparison. For a full list of parameters used, see Appendix \ref{app:parameters}.}

    \label{fig:layers}
\end{figure*}

By minimizing the energy in Eq. \ref{eqn:energy_cim}, the charge configuration of the system can be determined as a function of the applied voltages and a CSD is generated. An example of a simulated CSD for a DQD is shown in Fig. \ref{fig:layers} (a). In this Figure, the voltage space is divided into distinct honeycomb regions, each corresponding to a stable charge occupancy of the QDs. At low gate voltages, the system remains unoccupied, as the applied electrostatic potential is insufficient. As the gate voltage on one or both QDs increases, discrete charge transitions occur, leading to a stepwise increase in electron occupation. The boundaries between these regions correspond to charge addition lines, where the electrostatic potential of the QDs aligns with the Fermi level of the reservoir, allowing electrons to tunnel in, resulting in the observed honeycomb pattern.

The CIM is capable of generating basic simulations of CSDs; however, its current framework is limited to capturing the regular honeycomb pattern observed in Fig. \ref{fig:layers} (a). In the following section, we extend the CIM by introducing enhancements that account for additional features commonly seen in QD measurements. 

\section{Architecture of RF-Squad}
\label{sec:Simulation}

While the CIM offers an intuitive understanding of charge measurements in QD arrays, it falls short in capturing certain physical phenomena commonly observed in QD measurements. Examples of such phenomena include tunnel coupling, quantum confinement effects, and voltage dependencies of each element in the capacitance matrices of the CIM \cite{Guidoreview}. Moreover, it incorporates no knowledge of the tunneling rate between QDs and the reservoir, which is relevant in RF reflectometry experiments. To bridge this gap, we discuss the inclusion of additional layers of complexity in RF-Squad and their effects observed in a CSD, which cannot be captured using the CIM alone.

\subsection{Tunnel Coupling}

We first extend the CIM model by incorporating tunnel coupling, a quantum mechanical interaction that allows electrons to tunnel between neighbouring QDs. As demonstrated in Das Sarma et al. \cite{HubbardModel_CIM}, the CIM can be directly mapped to the spinless Hubbard Model, and a complete demonstration of this relationship for a DQD is provided in Appendix \ref{app:mapping_cim_hubbard}.

To include tunnel coupling into the simulations, we begin by constructing the Hamiltonian matrix for a QD array. We first need to define the basis states used to build the Hamiltonian, the Fock states. Each QD configuration can be represented by a Fock state, a sequence of numbers with length equal to the number of QDs, where each number represents the number of electrons occupying each QD. In mathematical terms, we consider an array of $M$ QDs, where the system can exist in one of $L$ possible Fock states, denoted as:

\begin{equation}
    \mathcal{F} = \left\{ \Lambda_j = (\lambda_{1,j}, \ldots, \lambda_{M,j}) \mid \lambda_{i,j} \in \mathbb{Z}^+, \, \forall j = 1, 2, \ldots, L \right\}, 
\end{equation}

\noindent where $\lambda_{i, j}$ represents the number of electrons in QD 
$i$ for a given Fock state $\Lambda_{j}$. These Fock states define the charge configurations of the system, forming the basis on which we construct the Hamiltonian.

In the absence of tunnel coupling, the Hamiltonian is diagonal, with each entry representing the energy of a specific charge configuration, as determined by the CIM. The inclusion of tunnel coupling introduces off-diagonal elements in the Hamiltonian, and the values of these terms represent the tunnel coupling strength between QDs. We denote these terms by $t_{p, q}$, which connect the Fock states where an electron hops between two QDs.

The resulting Hamiltonian matrix is

\begin{equation}
\resizebox{\columnwidth}{!}{$H = 
\begin{pmatrix}
\frac{1}{2}\,\vec{Q}_{1}^{\,T}\mathbf{C}_{dd}^{-1}\vec{Q}_{1} & t_{12} & t_{13} & \cdots & t_{1\ell} \\
t_{12} & \frac{1}{2}\,\vec{Q}_{2}^{\,T}\mathbf{C}_{dd}^{-1}\vec{Q}_{2} & t_{23} & \cdots & t_{2\ell} \\
t_{13} & t_{23} & \ddots & \ddots & \vdots \\
\vdots & \vdots & \ddots & \frac{1}{2}\,\vec{Q}_{\ell-1}^{\,T}\mathbf{C}_{dd}^{-1}\vec{Q}_{\ell-1} & t_{\ell-1,\ell} \\
t_{1\ell} & t_{2\ell} & \cdots & t_{\ell-1,\ell} & \frac{1}{2}\,\vec{Q}_{\ell}^{\,T}\mathbf{C}_{dd}^{-1}\vec{Q}_{\ell}
\end{pmatrix}
$}
\label{eqn:hamiltonian}
\end{equation}

\noindent where $\vec{Q}_j = e(\mathbf{C_{dg}}\vec{V_g} - \Lambda_j)$. Although all coupling terms are included in the Hamiltonian above, only the single-electron tunnel couplings between adjacent Fock states contribute appreciably, with all other terms being negligible by several orders of magnitude.

To determine the system's charge configuration at a set of applied voltages, we compute the lowest eigenvalue---which represents the ground state energy---and its corresponding eigenvector---which provides the electron distribution across the QDs. In Fig. \ref{fig:layers} (b), we illustrate this effect for a DQD system. Comparing panels (a) and (b), we observe that the inclusion of tunnel coupling introduces a smooth curvature along charge transition boundaries, where electrons tunnel between nearest neighbour QDs, with the amount of the bending being directly influenced by the magnitude of the tunnel coupling term. 

\subsection{Quantum Confinement Effects}

To model the effect of quantum confinement, we approximate the confinement potential of a QD as an anisotropic two-dimensional harmonic oscillator, a potential that is representative of the most common gate-defined QDs in semiconductor nanostructures. The corresponding energy eigenstates are known as Fock-Darwin states, which are added to the Coulomb interactions (from the CIM) in the total energy of the electron system~\cite{Kouwenhoven2001} 

\begin{equation}
    E(\vec{N}) = E_{\text{CI}}(\vec{N}) + \sum_{i=1}^M \hslash\omega_{xi}\left(n_i + \frac{1}{2}\right) + \hslash\omega_{yi}\left(m_i + \frac{1}{2}\right).
    \label{eqn:fd_energy}
\end{equation}

\noindent Here, $E_{\text{CI}}(\vec{N})$ represents the Coulomb interaction energy as described by the CIM, and the summation runs over all $M$ quantum dots in the system, with each dot contributing two confinement terms along the $x$ and $y$ directions. The quantities $n_i$ and $m_i$ represent the orbital quantum numbers of dot $i$, while $\omega_{xi}$ and $\omega_{yi}$ denote the corresponding confinement frequencies, which may differ ($\omega_{xi} \neq \omega_{yi}$) to account for potential anisotropy. Each eigenstate can further exhibit degeneracies, such as spin and valley degrees of freedom, leading to additional shell filling in silicon. Consequently, once an orbital level is filled with two electrons, extra energy (the confinement energy) is required to occupy the next state, breaking the 2D periodicity of the honeycomb pattern observed in panels (a) and (b). Figure~\ref{fig:layers}(c) illustrates this behaviour, where differently sized hexagonal regions emerge as a result.

\subsection{Voltage-Dependent Terms}

To enhance the simulator’s ability to generate realistic data, this subsection introduces two voltage-dependent modules in the: (i) tunnel coupling and (ii) the capacitance matrix. 

\subsubsection{Voltage-Dependent Tunnel Coupling in RF-Squad}

So far, we have assumed that all terms, including the tunnel coupling, are constants. However, in reality, terms such as tunnel coupling often depend on the applied gate voltages~\cite{laine2025}. In particular, as the voltage configuration changes, so does the potential barrier between QDs, directly affecting the wavefunction overlap, which determines the coupling between states. To model this effect, we apply the Wentzel-Kramers-Brillouin (WKB) Approximation, a semi-classical method for estimating quantum tunnel couplings. In this framework, the tunnel coupling $ t_{pq} $ is proportional to the magnitude of the wavefunction in the barrier region between the QDs and can be expressed as:

\begin{equation}
t_{p q} \propto \exp\left(-\frac{1}{h} \int_{x_1}^{x_2} \sqrt{2m^*(V(r) - E_c)} \, dr \right),
\label{eqn:WKB_full}
\end{equation}

 \noindent where $E_c $ is the charging energy for the electron transition, $ m^* $ is the effective mass of the electron, and $ x_1 $ and $ x_2 $ are the classical turning points, where $ V(r) = E_c $.

To further refine the voltage dependence of the tunnel coupling in RF-Squad, we express the potential $V(r) $ in Eq \ref{eqn:WKB_full}  using a Taylor expansion around a reference point. This allows us to approximate the tunnel coupling in a more tractable form:

\begin{equation}
t_{pq} = t_{pq,0} \exp \left( \mathbf{a} \cdot \mathbf{G} + b \right),
\label{eqn:wkb}
\end{equation}

 \noindent where $t_{pq,0} $ and $b $ are constants, $\mathbf{G} $ is the applied voltage vector and $\mathbf{a} $ is a vector determined by the Taylor expansion.

The effect of voltage-dependent tunnel coupling on CSDs is illustrated in Fig. \ref{fig:layers} (d). In the weak-coupling regime (bottom left of the CSD, referring to  values below 0.1 V on gates 1 and 2), the tunnel coupling is small and the honeycomb pattern remains sharp.
Transitioning to the intermediate coupling regime (middle of the CSD between values 0.1 - 0.2 V on gates 1 and 2), the tunnel coupling increases, introducing curvature to the charge transition lines. In the strong-coupling regime (top right of the CSD, with values above 0.2 V on gates 1 and 2), the two QDs effectively merge into a single QD, reflected by straight transition lines in the CSD.

While voltage-dependent tunnel coupling plays an important role in defining the interaction strength between QDs, another critical factor influencing CSDs is the voltage dependency of capacitance terms. In the next section, we account for capacitance changes with respect to gate voltage, further enhancing the simulations. 
 
\subsubsection{Voltage Dependent Capacitance Matrices}

Following the incorporation of voltage dependence into tunnel coupling, we now extend this consideration to the capacitance terms in the CIM. Unlike tunnel coupling, modelling voltage-dependent capacitance terms is significantly more complex, as it depends on multiple factors, including device geometry, material properties, and gate voltages applied, all of which influence the potential landscape \cite{Guidoreview}.

Finite-element solvers offer a rigorous approach to computing these dependencies but are computationally intensive, making them impractical for large-scale simulations \cite{QTCAD2025}. Experimentally, some studies (e.g., \cite{Rao:2024sqv}) have attempted to characterize these dependencies by fitting a linear function to mutual capacitances between QDs in the capacitance matrix. However, in general, the precise voltage dependence of capacitance terms remains largely unknown and poorly understood in QD physics.

To balance physical accuracy and computational efficiency, RF-Squad provides a flexible alternative: instead of treating the capacitance elements as fixed constants, users can define them as functions of the applied gate voltages. This approach introduces realistic voltage-dependent behaviour while maintaining efficient simulation times.

Fig. ~\ref{fig:layers}(h) illustrates the effect of this modification on the CSD, alongside the WKB approximation being applied to the tunnel coupling terms. In this example, we apply a linear dependence to each of the capacitance matrices' elements in terms of voltage, $C_{ij} = a_{ij} V_{g1} + b_{ij} V_{g2} + c_{ij}$ for both $\mathbf{C}_{dd}$ and $\mathbf{C}_{dg}$. The specific values for each coefficient used in the capacitance matrix are provided in Appendix \ref{app:parameters}. Initially, the expected regular hexagonal structure is observed in the lower-left region (approximately $V_{g2}=0.0 - 0.15V$). However, at higher gate voltages, a compression and shearing effect emerges, distorting the honeycomb structures, as well as the WKB approximation, further increasing the tunnel coupling. 

While voltage-dependent capacitance captures complex electrostatic effects, real experimental measurements rely on RF reflectometry to probe these charge transitions. To build a more realistic simulation, we now incorporate RF measurement techniques, allowing us to model how capacitance fluctuations manifest in experimentally measurable signals.

\subsection{Radiofrequency Reflectometry Simulations}

The previous layers of RF-Squad account for quantum mechanical effects that shape CSDs; they do not incorporate the measurement process itself. RF reflectometry is a widely used experimental technique for probing charge configuration in QDs, providing high-sensitivity detection of charge transitions. In this section, we focus on applying RF reflectometry in RF-Squad, allowing the simulation of realistic experimental readout \cite{Gonzalez-Zalba2015, West2018, vigneau2022, ciriano2021spin, Ibberson2021, urdampilleta2019gate, zheng2019rapid}.

When RF reflectometry is used, two methods for QD state readout exist: (i) charge-sensing, when an RF-driven sensor-QD detects changes in the electrostatic environment of nearest-neighbour QDs, examples of which are the RF single-electron transistor (SET) or the RF single-electron box (SEB). In this work, we primarily consider the latter~\cite{Oakes2023}. (ii) In-situ dispersive readout, when the RF is applied directly to the QD system to be measured and its polarizability is detected~\cite{Lundberg2020}.

The key measurable quantity in RF reflectometry is charge polarizability, which characterizes how the total charge of the system responds to small oscillations in an applied gate voltage. In an adiabatic approximation, this results in the differential capacitance \cite{Mizuta2016QuantumAT,Peri2024}:

\begin{equation}
    \Delta C_j = \frac{\text{d} Q_T}{\text{d} V_j},
    \label{eqn:C_j}
\end{equation}

\noindent where $ \Delta C_j $ is the differential capacitance, $ Q_T $ is the total charge on the sensing gate, and $ V_j $ is the oscillating voltage applied from gate $ j $. Since capacitance changes correspond directly to shifts in the reflected RF signal, measuring $ \Delta C_j $ provides an indirect method to probe electron transitions between QDs and reservoirs.

To compute $ \text{d} Q_T $, we consider both the total charge in each QD, $ -e \langle n_i \rangle $, and the effects of induced charge from gate electrodes as well as mutual electrostatic interactions between QDs.

For an $ M $-dot system, the total charge response to an applied gate voltage $ V_j $ is given by:

\begin{equation}
    \Delta C_{j} = \sum_{i=1}^M \text{d}C_{i, j} = e \sum_{i=1}^M  \alpha_{i, j} \frac{\text{d}\langle n_i \rangle}{\text{d}V_j} 
    \label{eqn:BIG}
\end{equation}

\noindent where $\alpha_{i,j}$ is an element of the lever-arm matrix $\mathbf{\hat{\alpha}}$, defined as:

\begin{equation}
    \mathbf{\hat{\alpha}} = \mathbf{C_{dd}^{-1}}\mathbf{C_{dg}}.
\end{equation}

\noindent This matrix links gate-induced potential changes to QD charge states, describing how applied voltages control electron occupancy.

To simulate RF capacitance changes, we first compute $ \langle n_i \rangle $, the average occupancy of QD $ i $, determined by the probability of the system occupying each possible Fock state:

\begin{equation}
    \langle n_i \rangle = \sum_{k = 1}^{L} \lambda_{i, k} \, P_{k},
    \label{eqn:n_i2}
\end{equation}

\noindent where $ \lambda_{i,k} $ represents the number of electrons in QD $ i $ for Fock state $ \Lambda_k $, $L$ is the maximum charge occupancy considered in the system, and $ P_k $ is the probability of the system being in $ \Lambda_k $.

Using the Boltzmann distribution, we express the probability of occupying state $ k $ as:

\begin{equation}
    P_{k} = \frac{\exp \left( -\frac{\epsilon_{k}}{k_bT} \right)}{\sum_{l=1}^L \exp \left( -\frac{\epsilon_{l}}{k_bT} \right)},
\end{equation}

\noindent where $ \epsilon_k $ is the energy of Fock state $ \Lambda_k $, $ k_b $ is the Boltzmann constant, and $ T $ is the system temperature.

By substituting $ P_k $ into Eq~\ref{eqn:n_i2}, we obtain:

\begin{equation}
    \langle n_i \rangle = \sum_{k = 1}^{L} \lambda_{i, k}  \frac{\exp \left( -\frac{\epsilon_{k}}{k_bT} \right)}{\sum_{l=1}^L \exp \left( -\frac{\epsilon_{l}}{k_bT} \right)}.
\end{equation}

In cases where electron wavefunctions extend across multiple QDs, the occupancy number $ \lambda_{i,k} $ can no longer be treated as an integer. Instead, we introduce a weighting factor $ p_i $ to capture the probability of an electron being delocalized across QDs:

\begin{equation}
    \langle n_i \rangle = \sum_{k = 1}^{L} p_i \, \lambda_{i, k} \,  \frac{\exp \left( -\frac{\epsilon_{k}}{k_bT} \right)}{\sum_{l=1}^L \exp \left( -\frac{\epsilon_{l}}{k_bT} \right)}.
\end{equation}

\noindent Here, $ p_i $ is derived from the square of the eigenvectors obtained when solving the system Hamiltonian, ensuring that non-localized charge distributions are properly accounted for in the simulation.

Using this formulation, RF-Squad computes the RF signal by evaluating the capacitance response given in Eq~\ref{eqn:BIG}. The total capacitance variation is:

\begin{align}
    \Delta C_{j, \text{tot}} &= \sum_{i=1}^M \text{d}C_{i,j} \\
    &= e \sum_{i} \alpha_{i,j} \frac{\text{d}}{\text{d}V_j} 
    \left(
    \sum_{k = 1}^{L} \lambda_{i, k} \,  \frac{\exp \left( -\frac{\epsilon_{k}}{k_bT} \right)}{\sum_{l=1}^L \exp \left( -\frac{\epsilon_{l}}{k_bT} \right)}
    \right).
\end{align}

\noindent By selecting the appropriate gate voltage for readout, RF-Squad simulates realistic RF measurements, capturing the dependence of the reflected signal on electron occupancy dynamics.

Fig.~\ref{fig:layers}(f) illustrates an example of this simulation, showing how the sensor dot (SD) detects charge transitions in a DQD system. Specifically, instead of using a charge sensor, we probe the DQD directly by applying the ac signal to the first QD. This results in the transitions where an electron transitions to the first QD, having a high differential capacitance because the probing signal is more sensitive to the first QD. Meanwhile, the electron transitions of the nearest neighbour QDs remain less pronounced due to the smaller lever arm between the probing gate and the QD.

\subsection{Tunnel Rates}

The next layer of RF-Squad aims to capture the implications of finite tunnel rates in an RF measurement, specifically the effect of tunnel rates between the QDs in the array and the QDs and external reservoirs. To accurately simulate these effects, it is essential to first identify the type of transition. Specifically, this involves distinguishing between an electron entering or leaving the QD system through an external reservoir (Dot to Reservoir Transition, DRT) and an electron tunnelling between two QDs (Interdot Charge Transition, ICT). Each type of transition is associated with a distinct tunnel rate and a corresponding probability of occurring.

To incorporate tunnel rates into RF-Squad, the probability of each transition for a given voltage configuration is calculated. This is determined using the Fermi function, which is defined as:

\begin{equation}
    \Gamma_{ij} = \frac{1}{1 + e^{\frac{E_i - E_j}{k_b T}}},
\end{equation}

where $ E_i $ and $ E_j $ are two eigenenergies of the QD system, obtained when solving the Hamiltonian outlined in Eq \ref{eqn:hamiltonian}. This function provides the likelihood of transitioning from an initial state with energy $ E_i $ to a final state with energy $ E_j $, based on thermal excitation.

The calculated transition probability is then multiplied by the thermal probability of the QD array occupying the initial state, given as $ e^{\frac{E_i}{k_b T}} $, where $ E_i $ represents the lowest energy eigenstate in a given pair. This results in the total probability of a transition, expressed as:

\begin{equation}
    P_{ij} = \Gamma_{ij} e^{\frac{E_i}{k_b T}}.
\end{equation}

\noindent Once the transition probabilities are determined, the type of transition, ICT or DRT, is identified by comparing the occupancy numbers of the QD system before and after the transition. Each transition type is assigned a specific weighting factor based on the tunnel rate and the probing frequency of the RF signal.

The overall RF measurement is then modified to account for tunnel rates using a weighted multiplication factor, as shown below:

\begin{equation}
\resizebox{\columnwidth}{!}{$
\begin{aligned}
\Delta C_{j, \text{tot}} 
&= 
e \sum_i \alpha_{i,j} 
\left( 
    P_{\text{DRT}} \frac{\gamma^2}{\gamma^2 + \hbar^2 \omega_f^2} 
    + P_{\text{ICT}} \frac{t_c^2}{t_c^2 + \hbar^2 \omega_f^2}
\right) 
\frac{dP_i}{dV_j}
\end{aligned}
$}
\label{eq:deltaC}
\end{equation}

\noindent where $ \gamma $ and $ \omega_f $ are the tunnel rate and RF resonator frequency, respectively, and $ t_c $ represents the tunnel coupling strength. $ P_{\text{DRT}} $ and $ P_{\text{ICT}} $ are the cumulative probabilities of DRT and ICT transitions occurring \cite{vigneau2022}.

\begin{figure}[h]
    \centering
    \includegraphics[width=1.1\linewidth]{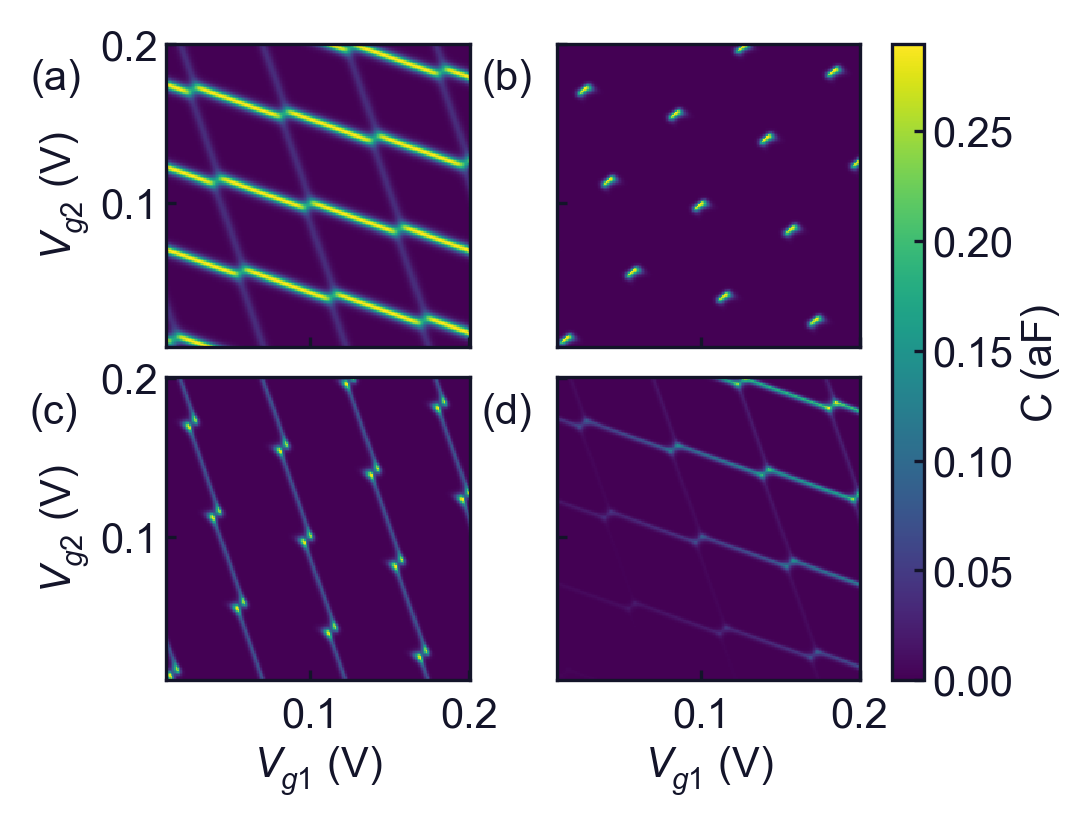}
    \caption{
        Effect of tunnel-rate weighting on the CSD of a DQD.
        (a) Baseline without tunnel-rate effects. 
        (b) ICT-dominated regime: $t_c \sim \hbar \omega_f$ and $\gamma_{\text{dot}} \ll \hbar \omega_f$, showing only interdot transitions. 
        (c) DRT-dominated regime (dot 1): $\gamma_{\text{dot1}} \sim \hbar \omega_f$ and $t_c \ll \hbar \omega_f$, yielding only dot\,1 reservoir features. 
        (d) Voltage-dependent tunnel rates (WKB): both ICT and DRT tunnel rates vary exponentially with gate voltages, as given by Eq.~\ref{eqn:wkb}. 
        For a full list of parameters used, see Appendix \ref{app:parameters}.
    }
    \label{fig:tunnel_rates_examples}
\end{figure}

This weighting has a simple consequence: transitions whose characteristic rate is comparable to the RF resonator frequency contribute strongly to the RF response, while much slower processes are suppressed. Figure~\ref{fig:tunnel_rates_examples} illustrates this behaviour. Panel~(a) shows the reference CSD without any tunnel-rate weighting. In panel~(b), the interdot coupling $t_c$ is chosen comparable to $\hbar \omega_f$, while the dot, reservoir tunnel rates $\gamma$ are negligible; consequently, only the ICT lines remain visible. In panel~(c) $t_c$ is made negligible and the tunnel rate for a dot to reservoir transition is set comparable to $\hbar \omega_f$. 
In (d), both $t_c$ and $\gamma$ are defined as voltage-dependent through the WKB approximation given in Eq.~\ref{eqn:wkb}, causing the tunnel rates to vary exponentially with the gate potentials. As a result, transitions only appear in regions with large applied gate voltages for gate 1 and gate 2. 

\subsection{Noise Models}

CSD measurements are inherently affected by various sources of noise, which must be accounted for to produce realistic simulations. In real experiments, background noise arises from electronic components, charge fluctuators, and material disorder in the QD environment \cite{Ferguson2006, Yoneda2017, spence2022, rudolph2019}. Two dominant sources of noise in these measurements are Gaussian noise and $1/f$ noise, both of which can be incorporated into RF-Squad as a post-processing layer, following the simulation of a CSD.

The first type of noise we focus on is Gaussian noise, which is usually dominated by the first amplification stage of the RF setup, usually a cryo-HEMPT, and is characterized by a normal distribution, given as:

\begin{equation}
    N_{\text{Gaussian}} = \frac{1}{\sqrt{2\pi\sigma^2}} \exp\left(-\frac{\mu^2}{2\sigma^2}\right),
\end{equation}

\noindent where $\mu$ is the mean of the noise, and $\sigma^2$ represents its variance, determining the spread of the distribution.

Another major noise source in QD measurements is $1/f$ noise, usually as a result of charge noise, which has a frequency-dependent spectral density and is given by:

\begin{equation}
    S_{1/f}(f) = \frac{A}{f^\alpha},
\end{equation}

\noindent where $S_{1/f}(f)$ is the power spectral density, $A$ is a scaling factor, $f$ is the frequency, and $\alpha$ is an exponent controlling the noise type. Setting $\alpha = 0, 1,$ or $2$ results in white, pink, or brown noise, respectively, while other values allow for intermediate noise types. In RF-Squad, $1/f$ noise is generated in the frequency domain, scaled according to $1/f^\alpha$, and transformed into the time domain using an inverse Fast Fourier Transform (IFFT), producing realistic low-frequency fluctuations observed in experiments.

These noise models are applied as an additional layer on top of the CSD simulation. Users can selectively enable or disable noise components and even stack multiple noise sources to replicate experimental conditions. An example of simulated background noise, specifically $1/f$ noise, is shown in Fig.~\ref{fig:layers} (g), where background fluctuations manifest as streaking artefacts, making transitions harder to detect as highlighted by  comparing (f) and (g). 

By incorporating tunable noise models, RF-Squad enhances the realism of charge stability simulations, allowing for more accurate comparisons between theoretical predictions and experimental data.

\section{Adaptive Fock States}
\label{sec:reduced_fock_states}

Simulating QD systems with multiple charge carriers requires solving the eigenvalue problem of a Hamiltonian whose dimension scales exponentially with both the number of QDs ($M$) and the maximum occupancy per dot ($N$). The Hamiltonian is constructed over a full Fock-state basis, where each basis vector represents a distinct charge configuration across the array. Consequently, the total Hilbert-space dimension grows as $(N + 1)^M$, rapidly increasing the computational cost of diagonalisation.  

To mitigate the increasing runtime, we expand upon an optimization algorithm, first introduced in \cite{qarray}, that selectively reduces the number of Fock states considered while maintaining accuracy.

This reduction is achieved by first treating the Fock states as continuous rather than discrete variables. By minimizing the energy expression from the CIM, as given in Eq \ref{eqn:energy_cim}, we identify a set of non-integer Fock states that correspond to the lowest-energy charge configuration for a given applied voltage:

\begin{equation}
\min_{\vec{N}}  \frac{1}{2}e^2 (\mathbf{C}_{\text{dg}} \vec{V}_g - \vec{N})^T \mathbf{C}_{\text{dd}}^{-1} (\mathbf{C}_{\text{dg}} \vec{V}_g - \vec{N}) ,
\end{equation}

\begin{equation*}
\vec{N} \in \mathbb{R}^n, \quad \vec{N} \geq 0.
\end{equation*}

To construct a reduced Fock basis, we round these continuous values up and down for each QD, selecting only the most probable states. This significantly decreases the number of Fock states used in the Hamiltonian while ensuring that the most relevant charge configurations are retained.

However, reducing the number of Fock states introduces a trade-off: while it speeds up computation, it can also introduce artefacts in CSDs (see Appendix \ref{app:artefacts} for examples). This effect is particularly pronounced in simulations with large tunnel coupling terms or voltage-dependent capacitance matrices, where excluding additional Fock states may alter charge transition boundaries. To counteract this, RF-Squad introduces the Maximum Fock Depth (MFD) parameter, which controls the number of additional Fock states included in the Hamiltonian by incorporating a greater number of values both above and below the continuous values obtained from the minimization calculation. Increasing MFD improves accuracy by capturing more relevant charge configurations, but comes at the cost of longer simulation times.

\begin{figure}[h!]
    \centering
    \includegraphics[width=\linewidth]{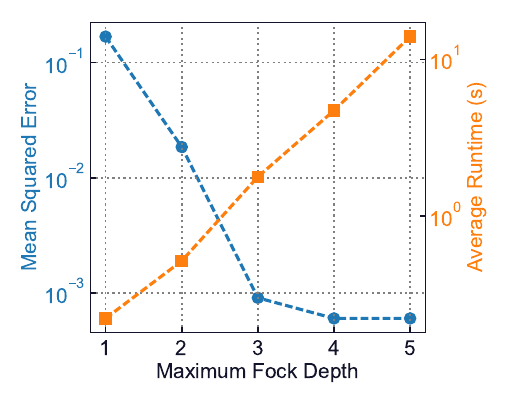}
    \caption{Dependence of the mean squared error (MSE) on the maximum Fock depth in the reduced Fock basis approach. Increasing the maximum Fock depth systematically improves simulation accuracy by reducing artefacts, but at the cost of longer computation times. Each point corresponds to a 100$\times$100-pixel CSD simulation performed using the full Fock basis, including voltage-dependent tunnel couplings and quantum confinement effects. The corresponding reduced simulations were performed with adaptive Fock depths (MFD = 1–5), and the MSE was computed relative to the full Fock state reference to quantify the resulting error efficiency trade-off.}
    \label{fig:time_comparison_MSE}
\end{figure}

Fig. \ref{fig:time_comparison_MSE} illustrates this trade-off by comparing the runtime and mean squared error of reduced Fock state simulations at different values of MFD against the full Fock basis calculation. Here, we simulated a 100 x 100 pixel CSD of a DQD containing features comprising voltage-dependent tunnel coupling and quantum confinement effects. All parameters can be found in Appendix \ref{app:parameters}. 
As shown, reducing the number of Fock states significantly lowers computation time, particularly as the number of charge carriers increases. Notably, beyond a certain MFD value, additional charge states contribute little to accuracy improvements, suggesting an optimal balance between speed and precision.

By implementing this reduced Fock state approach, RF-Squad maintains computational efficiency even in simulations involving large charge carriers while allowing users to fine-tune accuracy based on their specific needs. This method ensures that while full Fock state calculations remain computationally expensive, a well-chosen MFD can achieve near-identical results with significantly lower runtime.

\section{Benchmarking and Performance}
\label{sec:benchmarking}

In the previous section, we explored how RF-Squad extends beyond the CIM to generate more realistic CSDs. Next, we focus on RF-Squad's computational performance. We benchmark its performance by measuring simulation time as a function of key parameters: the total number of QDs, charge carriers, and simulation layers. All tests were conducted on a MacBook Air (2023) with an Apple M2 chip, 16 GB of RAM, running macOS Sequoia 15.1. For each of the times stated in this section, the average time was taken of the total number of devices simulated and the standard error of the mean was used to calculate the error. 

\subsection{Number of Quantum Dots}

We begin by analyzing the computational performance of RF-Squad as a function of the number of QDs. 
In this context, increasing the number of QDs corresponds to expanding the capacitance matrix to an 
$M \times M$, where $M$ is the number of QDs. For each value of $M$, we considered systems with an 
equal number of gates and QDs. To generate the CSDs, two gate voltages 
(chosen from the $M$ available gates) were swept simultaneously while the remaining gate voltages 
were held fixed. For each configuration, we averaged the runtime over repeated independent simulations. The maximum occupancy per QD was varied from one to three. 

\begin{figure}[h!]
    \centering
    \includegraphics[width=\linewidth]{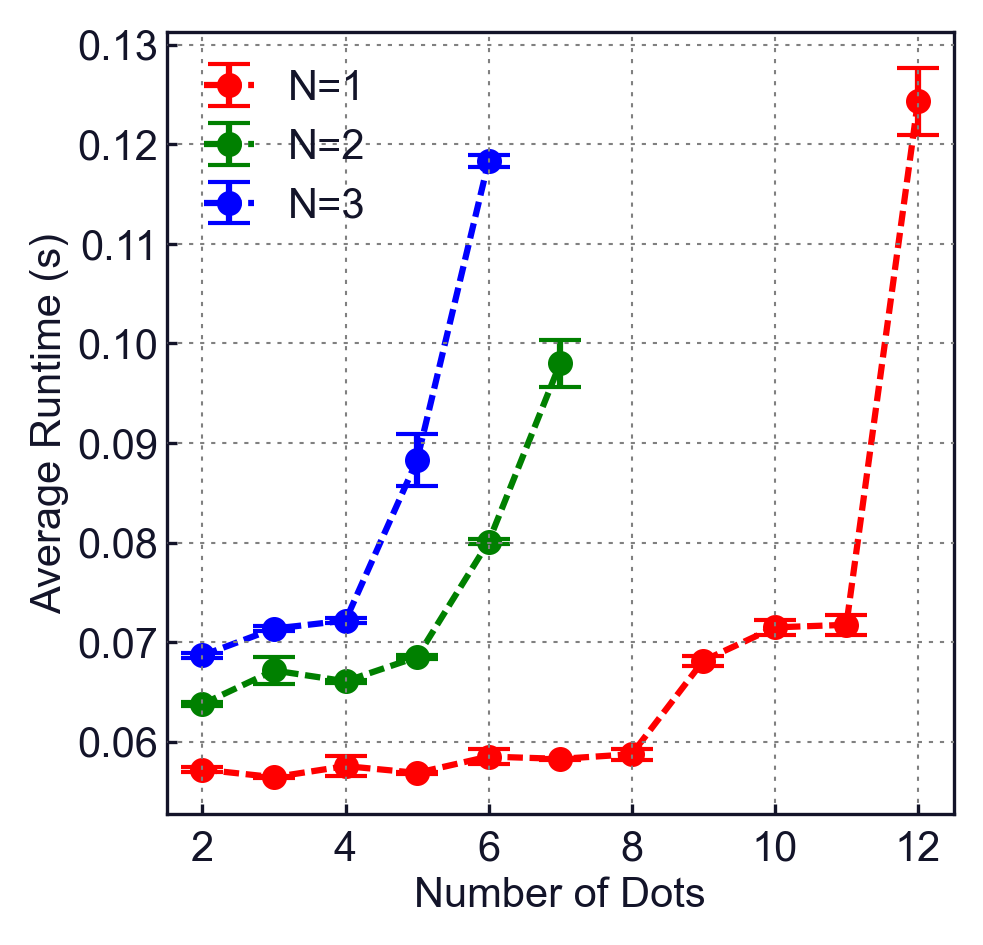}
    \caption{The average time required to run the base layer of RF-Squad, the CIM, as a function of the number of QDs. The average was determined by simulating the occupancy diagrams with $100 \times 100$ pixels for one hundred random QD arrays ranging in size by increasing the size of the capacitance matrix $M$. The maximum QD occupancy ranged from one to three.}
    \label{fig:num_dots_v_time}
\end{figure}

To benchmark performance, we generated 100$\times$100 pixel CSDs for one hundred randomly generated QD arrays, increasing the value of $M$, and thereby increasing the number of QDs in the array. To isolate the base computational cost, we used only the CIM layer (as discussed in Section \ref{sec:CIM}), which requires only matrix multiplication. The results, shown in Fig. \ref{fig:num_dots_v_time}, indicate that for a DQD with maximum occupancy of one, the average runtime was approximately 52.1 $\pm 0.2$ milliseconds. As the number of QDs increased, the computational time grew, reaching $124.3 \pm 0.4 ~\mathrm{ms} $ milliseconds for a system with twelve QDs using a maximum occupancy of one charge carrier per dot. 

Increasing the maximum occupancy per QD had a more pronounced effect on the computational cost.  
For a double quantum dot (DQD, \(M = 2\)), increasing the maximum electron occupancy per dot from one to two and three resulted in average runtimes of \(63.8~\mathrm{ms}\) and \(68.7~\mathrm{ms}\), respectively. Extending the system size further, we obtained runtimes of \(99 \pm 1~\mathrm{ms}\) for a six-dot device with \(N = 2\), and \(118 \pm 2~\mathrm{ms}\) for a seven-dot device with \(N = 3\). We note that for larger arrays the runtime grows substantially. This increase is primarily due to the rapidly expanding memory footprint, which causes the solver to exceed fast cache capacity and rely on slower backed-up memory access.

\noindent Figure~\ref{fig:num_dots_v_time} highlights this behaviour, showing a growth in computation time for increasing $M $, with minor deviations arising from random variations in the capacitance matrices.  
Importantly, the results demonstrate that even for moderately large QD systems ($M = 12 $), simulations remain computationally tractable on standard hardware, with runtimes well below a second per CSD.  

These findings establish the baseline performance of the CIM layer and confirm its scalability for multi-dot arrays. In the following section, we extend this analysis how the RFS method mitigates their computational overhead.  

\subsection{Maximum Quantum Dot Occupancy}

Beyond the number of QDs, the computational efficiency of RF-Squad is also heavily influenced by the maximum number of charge carriers each QD can hold. Each additional charge carrier increases the number of possible Fock states, expanding the size of the Hamiltonian matrix and increasing the computational complexity of the simulation.

\begin{figure}[h!]
    \centering
    \includegraphics[width=\linewidth]{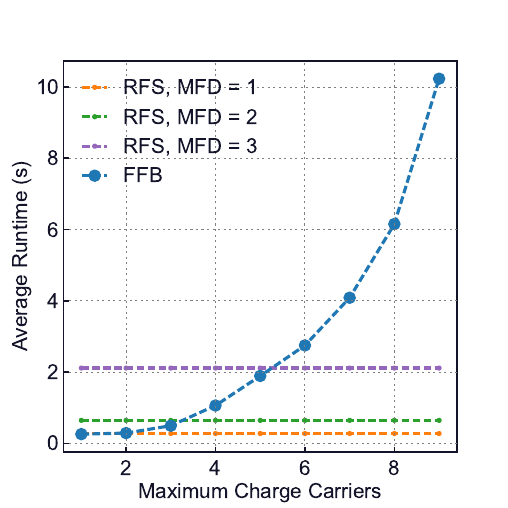}
    \caption{Comparison of average run times for generating occupancy diagrams as a function of the maximum number of charge carriers in QD arrays. The reduced Fock state algorithm with varying maximum Fock depths (MFD = 1, 2, 3) and full Fock basis (FFB) is shown. Simulations were performed on 100$\times$100 pixel occupancy diagrams for one hundred random QD arrays, with the number of gates equal to the number of QDs with varying maximum number of charge carriers.}
    \label{fig:time_comparison_2}
\end{figure}

To quantify this effect, we benchmarked RF-Squad’s performance as a function of maximum charge carriers per QD, measuring runtime while varying the occupancy limit.  CSDs were simulated for one hundred random DQDs as a function of the maximum number of charge carriers per QD.  Both the RF and tunnel coupling layers were included in this simulation, as the RFS would most likely be used in simulations beyond the CIM layer. Each simulation generated a CSD of 100$\times$100 pixels. The number of gates was set to the number of dots, and in each simulation, both gate voltages were swept. These results, presented in Fig. \ref{fig:time_comparison_2}, illustrate how increasing charge occupancy affects simulation speed and highlight optimization strategies to reduce computational overhead.

The blue line in Fig.~\ref{fig:time_comparison_2} quantifies the computational scaling introduced in Sec.~\ref{sec:reduced_fock_states}. For a DQD, the average simulation time increases from $52.1 \pm 0.2~\mathrm{ms}$ for a maximum occupancy per QD of one, to $10.23 \pm 0.03~\mathrm{s}$ for an occupancy of nine. These measurements highlight the practical motivation for introducing the reduced Fock-state method, which significantly mitigates this computational overhead while preserving spectral accuracy.

The remaining curves in Fig.~\ref{fig:time_comparison_2} correspond to simulations performed using the reduced Fock state approach, where the MFD is limited to 1, 2, or 3. In this case, the runtime remains independent of the number of charge carriers, since the number of Fock states, and therefore the size of the Hamiltonian, is fixed by the chosen Fock depth rather than the occupancy limit. As expected, increasing the MFD leads to longer runtimes, reflecting the larger basis used in the calculation. However, as discussed previously, this increase in computational cost is offset by a reduction in the accuracy of certain higher-order charge configurations, which are truncated when using the reduced Fock state representation.

Charge occupancy is not the only factor affecting computational efficiency. The complexity of the simulation also depends on the number of physical effects included, each discussed in Section \ref{sec:Simulation}. In the next section, we explore how these additional layers influence both computational cost and the fidelity of simulated charge stability diagrams.

\subsection{Simulation Complexity}

Beyond QD occupancy, another key factor influencing RF-Squad’s performance is the complexity of the physical model used in a simulation. 
RF-Squad’s layered architecture allows users to select the desired level of physical detail, from simple electrostatic models to fully quantum-mechanical simulations. 
While greater complexity naturally demands more computation, our benchmarking shows that once a simulation moves beyond the basic CIM, 
additional layers of physical realism provide substantial accuracy gains for only a modest increase in computational cost.

\begin{figure}[h!]
    \centering
    \includegraphics[width=\linewidth]{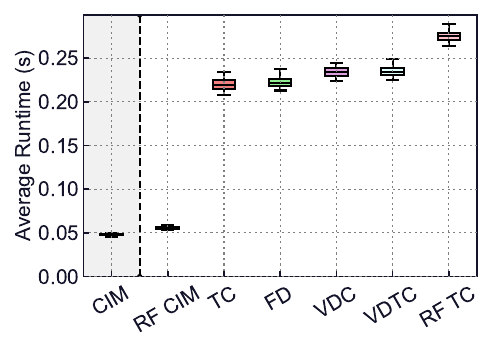}
    \caption{    Time required to simulate a CSD of $100 \times 100$ pixels for one hundred random DQD devices, each with a maximum charge-carrier occupancy of one. 
    The results are shown for each simulation layer within RF-Squad: the constant interaction model (CIM), tunnel coupling (TC), quantum confinement (QC), voltage-dependent tunnel coupling (VDTC), and voltage-dependent capacitances (VDC). 
    RF-CIM and RF-TC denote configurations that include RF reflectometry readout with the CIM and TC models, respectively.
    We highlight the CIM as the foundational layer, upon which all subsequent physical models are constructed; however, each layer can also be executed independently for benchmarking and modular testing.
    The maximum QD occupancy was fixed to one.}
    \label{fig:time_comparison}
\end{figure}

To quantify this, we measured the time required to simulate a CSD of $100 \times 100$ pixels for one hundred randomly generated DQD arrays, with the maximum electron occupancy per QD set to one. Each layer of RF-Squad was benchmarked independently, and the average results are shown in Fig.~\ref{fig:time_comparison}.

As expected, the CIM is the fastest, as it involves only basic matrix operations. 
The largest computational jump occurs when moving from the CIM to models requiring full Hamiltonian diagonalisation. 
Including (i) tunnel coupling, (ii) quantum confinement, (iii) voltage-dependent tunnel coupling, or (iv) voltage-dependent capacitance matrices yields average runtimes of $224.6 \pm 4.5~\mathrm{ms}$, $224.0 \pm 1.1~\mathrm{ms}$, $235.8 \pm 1.6~\mathrm{ms}$, and $235.2 \pm 1.2~\mathrm{ms}$, respectively. 
Once this transition is made, further quantum effects increase the cost by at most about 5\%, 
while providing a more physically accurate description of the system. By comparison, a full RF simulation with tunnel coupling takes $280.8 \pm 4.3~\mathrm{ms}$, representing a clear but expected additional overhead (about 19\%).

Interestingly, incorporating RF measurements, whether applied to the CIM or to more complex models, adds only a small runtime overhead. 
This suggests that once users employ a fully quantum-mechanical model, enabling all available physical layers offers the most accurate charge stability simulations with minimal additional computational cost.

\section{Examples}
\label{sec:examples}
Having established RF-Squad's computational performance in benchmarking tests, we now demonstrate its practical applications in simulating CSDs and replicating experimental tuning procedures. These examples highlight RF-Squad’s ability to serve as both a data generation tool and a digital twin for QD experiments.  
In the following subsections, we explore specific examples that illustrate these capabilities. For information on the parameters used or the mean squared error between simulated and experimental measurements, see Appendix \ref{app:parameters}.

\subsection{Sensor Dot and Double Quantum Dot}

To illustrate RF-Squad’s ability to model realistic QD behaviour, we first compare an experimental measurement of a DQD with a SD to a corresponding RF-Squad simulation.

\begin{figure*}
\centering
    \includegraphics[width=0.98\textwidth]{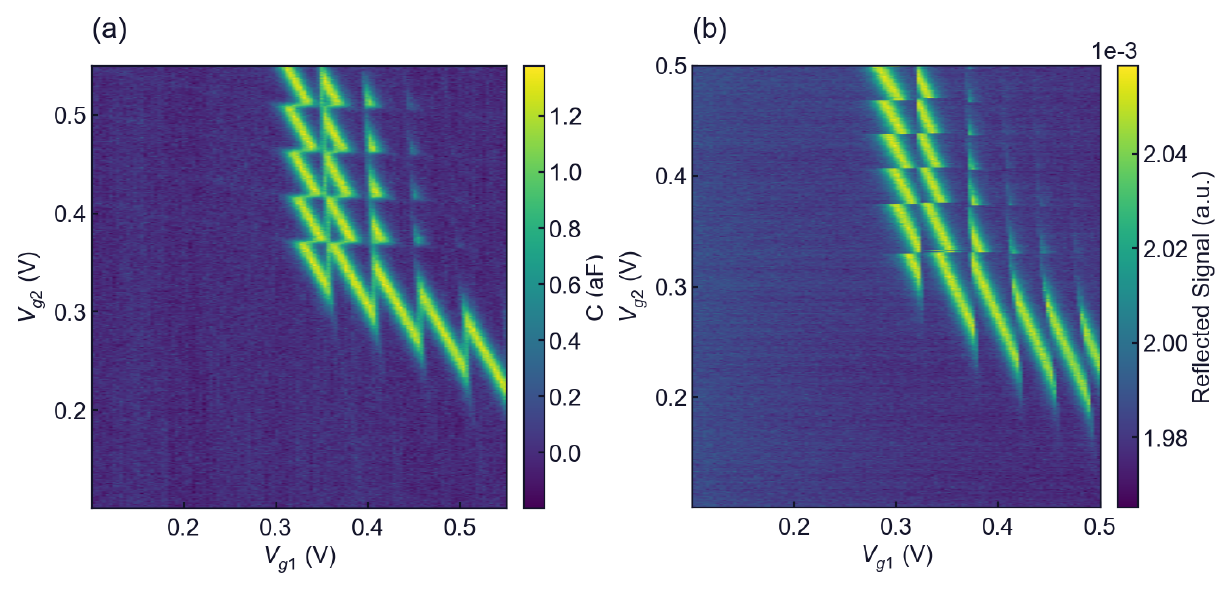}
    \caption{(a) Simulated data of a DQD with a SD. Here, the two applied gates to the DQD are swept, and the SD is used to measure the CSD. (b) Experimental data of a DQD measurement using an adjacent SD in RF. For information about the experiment of the parameters used for the simulation, see Appendix \ref{app:parameters}.}
    \label{fig:SET_simulation}
    
\end{figure*}

The system consists of three QDs: two forming a DQD, while the third serves as a SD to detect charge transitions via RF reflectometry, in this case, a single-electron box. 
The simulated RF reflectometry map (Fig.~\ref{fig:SET_simulation}a) is compared with experimental data (Fig.~\ref{fig:SET_simulation}b).
In both the experiment and simulation, we sweep the gate voltages controlling the DQD while keeping the SD voltage and barrier gate between the QDs at a fixed voltage. For a full list of parameters used, see Appendix \ref{app:parameters}. 

By directly comparing the simulated and experimental CSDs, one can see that it visually captures the key properties of a CSD, demonstrating strong agreement between the simulated and experimental data. 

Although this agreement is promising, CSDs are also influenced by the dynamics of electron tunnelling, which affect how charge states evolve over time. To explore this effect, we next examine how tunnel rates impact CSDs and determine how RF-Squad can replicate these time-dependent properties.

\subsection{Impact of Tunnel Rates on in-situ DQD RF measurements}

\begin{figure*}
    \centering
    \includegraphics[width=1.02\textwidth]{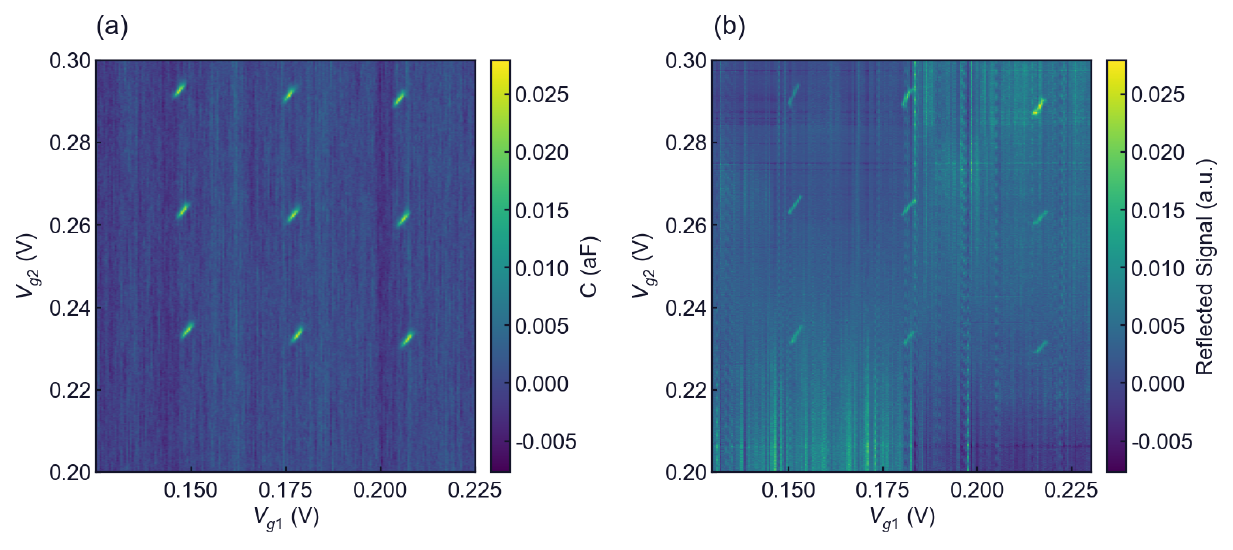}
    \caption{(a) Simulated data illustrating the effect of tunnel rates on the CSD, emphasizing the decrease of signal of the QD to reservoir (DRT) transitions and the amplification of ICTs under specific probing frequency. For the set of parameters used, see Appendix \ref{app:parameters}. (b) Experimental data of a DQD measured dispersively \textit{in-situ}.}
    \label{fig:tunnel_rates}
\end{figure*}

To further investigate the role of tunnel rates in QD behaviour, we simulate their effect on the CSD of a DQD system, as shown in Fig. \ref{fig:tunnel_rates}. The capacitance matrix used as input parameters, as well as the tunnel rates for this simulation, is provided in Appendix~\ref{app:parameters}.

In this scenario, the DQD is measured \textit{in-situ}, and the probing frequency is closely matched to the tunnel coupling between the QDs, resulting in a significant dispersive signal, leading to well-resolved ICTs \cite{Peri2025}. In contrast, the tunnel rates associated with the reservoirs are far from the probing frequency, resulting in the suppression of the DRTs~\cite{Peri2024SEB}. These effects are clearly visible in both the simulated and experimental CSDs, as illustrated in Fig. \ref{fig:tunnel_rates}.

\section{Conclusions}
\label{sec:conclusions}

In this work, we have presented RF-Squad, a physics-based simulator for RF measurements of QD arrays, designed to strike an optimal balance between physical accuracy and computational efficiency. Through its modular, layered architecture, RF-Squad enables users to progressively incorporate quantum mechanical effects, such as tunnel coupling, confinement, voltage-dependent parameters, and finite tunnel rates, while retaining rapid simulation times. Implemented in JAX, it achieves millisecond-scale generation of charge stability diagrams, providing a scalable foundation for digital-twin modelling of QD systems.

By benchmarking its performance and validating against experimental data, we have demonstrated that RF-Squad reproduces key experimental features of RF reflectometry, capturing both interdot and dot reservoir transitions with realistic noise and frequency dependent effects. Beyond its use as a simulation engine, RF-Squad offers a practical framework for generating large datasets to train and test machine learning algorithms for device autotuning.

Looking ahead, RF-Squad establishes a foundation for next generation QD simulation platforms, supporting not only the development of automated tuning and control strategies, but also advancing large scale quantum device design, characterization, and integration into emerging quantum computing architectures.

\section{Acknowledgements}

This research was supported by the European Union’s
Horizon 2020 research and innovation programme under
grant agreement no. 951852 (QLSI), and by the UK’s Engineering and Physical Sciences Research Council (EPSRC) via the Cambridge NanoDTC (EP/L015978/1).
M.F.G.Z. acknowledges a UKRI Future Leaders Fellowship [MR/V023284/1]. T.M. acknowledges funding from
the Winton Programme for the Physics of Sustainability.


%

\appendix

\section{Linking the Constant Interaction Model and the Hubbard Model}

\label{app:mapping_cim_hubbard}
Here we repeat the derivations from Das Sarma et. al \cite{HubbardModel_CIM} and show the direct relationship between the Generalized Hubbard Model (GHM) and the Constant Interaction Model (CIM). The Hamiltonian of the GHM is given as; 
\begin{equation}
H = H_\mu + H_U + H_t
\end{equation}
which is composed of the chemical potential term $ H_\mu $, the Coulomb interaction term $ H_U $, and the tunneling term $ H_t $. 

Let us define  $c_{i\sigma}^{\dagger}$ and $c_{i\sigma}$ as the creation and annihilation operators, respectively, for an electron of spin $\sigma$ at site $i$. We define the number operator as $n_{i\sigma} = c_{i\sigma}^{\dagger} c_{i\sigma}$ for electrons with spin $ \sigma $ on site $ i $ and $n_i = n_{i\uparrow} + n_{i\downarrow}$. Using these definitions, each term of the GHM can be written explicitly. The chemical potential term $ H_\mu $ is given by

\begin{equation}
H_\mu = \sum_{i} -\mu_i n_i, 
\end{equation}
which represents the single site terms of the Hamiltonian and $\mu_i$ being the chemical potential of dot $i$. The Coulomb interaction term $ H_U $ can be written as 

\begin{equation}
H_U = \sum_{i}  U_i n_{i\uparrow} n_{i\downarrow} + \sum_{i, j} U_{ij} n_i n_j.
\end{equation}
Here, $U_i$ represents the on-site Coulomb interaction energy and $U_{ij}$ represents the inter-site Coulomb interaction energy. The final term, which is the tunnelling term $ H_t $ can be written as 

\begin{equation}
H_{t} = - \sum_{\sigma} \sum_{i, j} t_{c} (c_{i\sigma}^{\dagger} c_{j\sigma} + c_{i\sigma}^{\dagger} c_{j\sigma}).
\end{equation}
Here,  $t_c$ represents the tunnelling or hopping term, describing the hopping of electrons between sites.

To simplify this model and relate it to the CIM, spin can be neglected, which would result in $n_{i\downarrow} = n_{i\uparrow} = N_i$. One can also assume that the tunnel coupling $t_c = 0$. This results in the  spinless Hubbard Model, which is defined as; 

\begin{equation}
    H = -\sum_{i} \mu_i N_i  +  \sum_{i}  U_i N_{i}^2 + \sum_{i, j} U_{ij} N_i N_j,
    \label{eqn:hubb2}
\end{equation}

A further assumption that can be used is that the electrochemical potential is directly proportional to the applied voltages, and is given by

\begin{equation}
\mu_i = |e| \sum_{j}(\alpha_{ij} V_{g,j} + \gamma_j),
\end{equation}

where $\alpha_{ij}$ represents the strength of the relationship between site $i$ and gate $V_{g,j}$ and $\gamma_j$ is a translation factor. 

The electrochemical potential $E$ of the quantum dot system as defined in the CIM \cite{Wiel2002} is given as:

\begin{equation}
    E = \frac{1}{2} \vec{V}^{T} \mathbf{C^{-1}_{dd}} \vec{V},
    \label{eqn:energy_cim2}
\end{equation}

where we define $\vec{V} = e (\mathbf{C_{dg}} \vec{V_g} - \vec{N})$, where $e$ is the electron charge, $\vec{N}$ signifies the number of charges on each dot, known as the Fock state, and $\vec{V_g}$ denotes the applied gate voltages.

As an example, we focus on a double quantum dot where $\vec{V_g} = (V_{g1}, V_{g2})$ and $\vec{N} = (N_{1}, N_{2})$. Inserting these in $\vec{V}$ and expanding the energy equation we obtain; 

\begin{equation}
E(N_1, N_2) = \frac{1}{2}N_1^2E_{C1} + \frac{1}{2}N_2^2E_{C2} + N_1N_2E_{Cm} + f(V_{g1}, V_{g2}),
\label{eqn:U(n1n2)}
\end{equation}

where we define $f(V_{g1}, V_{g2})$ as 

\begin{align}
f(V_{g1}, V_{g2}) = \frac{1}{|e|} \Big\{ & C_{g1}V_{g1}(N_1E_{C1} + N_2E_{Cm}) \notag \\
& + C_{g2}V_{g2}(N_1E_{Cm} + N_2E_{C2}) \notag \\
& + \frac{1}{e^2} \left( \frac{1}{2}C_{g1}^2V_{g1}^2E_{C1} + \frac{1}{2}C_{g2}^2V_{g2}^2E_{C2} \right) \notag \\
& + C_{g1}V_{g1}C_{g2}V_{g2}E_{Cm} \Big\}
\end{align}

and we denote the charging energies of dot one (two) as $E_{C1(2)}$ and the mutual energy $E_{C_m}$ as

\begin{equation}
E_{C1} = \frac{e^2}{C_1} \left( \frac{1}{1 - \frac{C_m^2}{C_1 C_2}} \right),
\end{equation}

\begin{equation}
E_{C2} = \frac{e^2}{C_2} \left( \frac{1}{1 - \frac{C_m^2}{C_1 C_2}} \right),
\end{equation}

\begin{equation}
E_{Cm} = \frac{e^2}{C_m} \left( \frac{1}{\frac{C_1 C_2}{C_m^2} - 1} \right).
\end{equation}

Above, we have assumed the cross capacitances (off-diagonal elements of $C_{dg}$) are negligible to simplify the equations, which is valid under the condition that the direct capacitive coupling between the quantum dots and the gates dominates over the cross-capacitive coupling terms. 

Here, $C_{1(2)}$ is the sum of all capacitances attached to dot $1(2)$, including $C_m$: 
$C_{1(2)} = C_{g1(2)} + C_m$. 

For a double quantum dot, the spinless Hubbard Model Hamiltonian is given as 

\begin{equation}
    H = \mu_1 N_1 + \mu_2 N_2 + U_1 N_{1}^2 + U_2 N_{2}^2  + U_{12} N_1 N_2,
    \label{eqn:hubbard_dqd}
\end{equation}

and the chemical potentials are given by; 

\begin{equation}
\mu_1 = |e| (\alpha_{11} V_1 + \alpha_{12} V_2) + \gamma_1,
\end{equation}

\begin{equation}
\mu_2 = |e| (\alpha_{21} V_1 + \alpha_{22} V_2) + \gamma_2.
\end{equation}

By comparing Equation \ref{eqn:U(n1n2)} for the electrostatic energy of a double quantum dot with the Hamiltonian of the Hubbard Model in Equation \ref{eqn:hubbard_dqd}, we can see that $U_i = E_{Ci}$ and $U_{12} = E_{Cm}$. Meanwhile, the proportionality terms $\alpha_{ij}$ can be grouped into a matrix and is mapped to $\hat{\alpha} = \mathbf{C_{dd}^{-1}} \mathbf{C_{dg}}$ while the terms $\gamma_i = -U_i/2$. 

While this has demonstrated a mapping between the CIM and Hubbard Model for a double quantum dot, it is important to note that such a direct mapping can be extended to any size $ M $ quantum dot device provided that the tunnel coupling is negligible. 

\section{Artefacts present in reduced Fock states}
\label{app:artefacts}

\begin{figure}[h!]
    \centering
    \includegraphics[width = \linewidth]{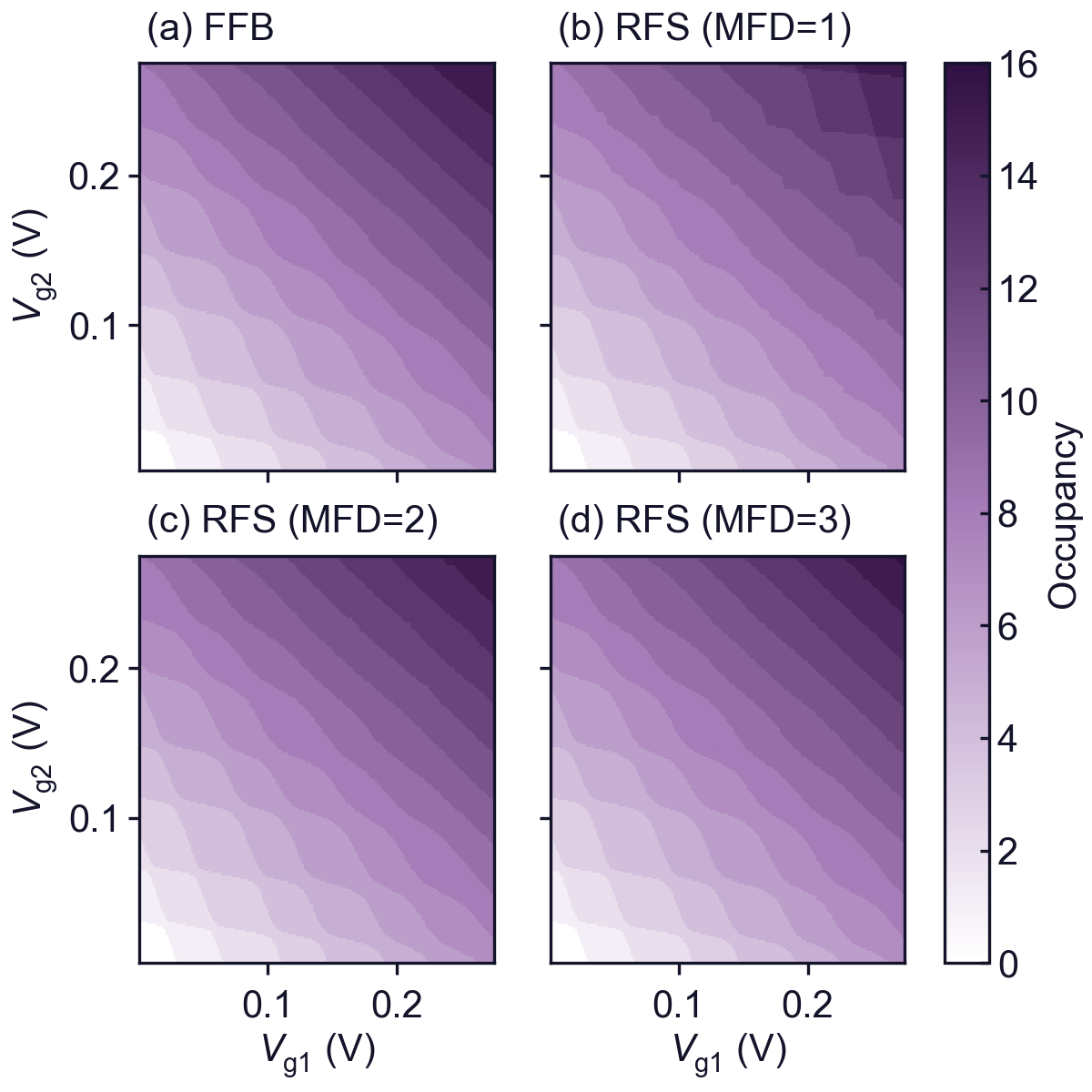}
    \caption{Occupancy diagrams for a double quantum dot illustrate the effect of increasing the maximum Fock depth on the mitigation of artefacts while using the reduced Fock state algorithm. Each panel corresponds to a specific maximum Fock depth, ranging from 1 to 5, and the full Fock basis (FFB). The progression demonstrates the disappearance of artefacts as the maximum Fock depth is increased, with the results converging toward the full Fock space implementation.}
    \label{fig:artefacts}
\end{figure}

Artefacts in the CSD can arise due to reduced Fock state representations, particularly in the presence of large tunnel couplings. As these off-diagonal elements of the Hamiltonian increase, the Fock states become strongly mixed, and the true eigenstates of the system transition from being pure Fock states to superpositions thereof. By truncating the Fock space, the reduced Fock state approach effectively omits part of this superposition, leading to distortions or artefacts in the simulated CSD. Increasing the maximum Fock depth progressively restores these missing components, thereby mitigating the artefacts and improving agreement with the full Fock basis results.  To eliminate these artefacts, we introduced the maximum Fock depth (MFD), which sets the number of Fock states included in the simulation. By progressively increasing the MFD, we observe that initially present artefacts gradually disappear, and the results converge toward the full Fock space implementation.
To highlight this effect, we explore the use fo the MFD model, which is shown in Figure \ref{fig:artefacts}.

When MFD = 1, a diamond-shaped pattern appears at large gate voltages where tunnel coupling is strong (see Figure \ref{fig:artefacts}(a) for voltages greater than 0.2 V), which is not seen in the FFB. This arises because a low MFD truncates essential Fock states that account for electron delocalization, which becomes significant as the double quantum dot transitions toward a single-dot regime. As the MFD increases (MFD = 2,3), these artefacts diminish, and for MFD = 4 or 5, they vanish entirely. At this point, the computed CSD becomes identical to the full-Fock-space calculation, which considers up to eleven charge carriers.

\section{Parameters used for Simulations}
\label{app:parameters}

This appendix summarizes the parameters used for all simulations presented throughout the manuscript.  
Unless otherwise stated, all simulations were performed using an electron temperature $T = 0.5~\mathrm{K}$, 
and capacitances expressed in attofarads (aF).  
Noise models and tunnel-rate conventions are consistent with those described in the main text.
\subsection*{Figure~\ref{fig:layers}}
A DQD device with two gates was simulated with
\[
C_{dd} =
\begin{pmatrix}
200 & -20 \\ -20 & 200
\end{pmatrix}~\mathrm{aF}, \qquad
C_{dg} =
\begin{pmatrix}
2.4 & 1.5 \\ 0.5 & 3.85
\end{pmatrix}~\mathrm{aF}.
\]
The tunnel coupling was $t_{12} = 0.5\times10^{-4}$~eV, and the confinement frequencies were 
$(\omega_{x,1},\omega_{y,1})=(3.0,4.0)\times10^{-4} \mathrm{eV}$ and 
$(\omega_{x,2},\omega_{y,2})=(5.0,6.0)\times10^{-4}\mathrm{eV}$.
Voltage-dependent tunnel rates followed Eq.~\ref{eqn:wkb} with $a=[0,15]$, $b=2$.
The gate–dot capacitance matrix $C_{dg}(V)$ was varied with the applied gate voltages according to
\[
C_{dg}(V) =
\begin{pmatrix}
C_{dg}^{(1,1)} & C_{dg}^{(1,2)}\,[1 - 4V_{g1}] \\
C_{dg}^{(2,1)}\, & C_{dg}^{(2,2)}
\end{pmatrix}~\mathrm{aF},
\]

Two noise sources were included: (i) $1/f$ noise with amplitude $A=10^{-3} \mathrm{W^2/Hz}$, and 
(ii) Gaussian noise with $\mu=0.05 ~\mathrm{aF}$, $\sigma=0.01 ~\mathrm{aF}$.

\subsection*{Figure~\ref{fig:tunnel_rates_examples}}
A DQD device with three gates was simulated with
\[
C_{dd} = 
\begin{pmatrix}
100 & -10 \\ -10 & 100
\end{pmatrix}~\mathrm{aF}, \qquad
C_{dg} =
\begin{pmatrix}
3.0 & 0.1 & 3.0 \\ 0.8 & 0.1 & 0.8
\end{pmatrix}~\mathrm{aF},
\]
where the middle column corresponds to the barrier gate.
Gate voltage ranges were $V_{\mathrm{g1}},V_{\mathrm{g3}}\!\in[0.01,0.20]$~V and 
$V_{\mathrm{g2}}\!\in[0.01,0.25]$~V.

\noindent
Parameter sets per panel:  
(a) no tunnel-rate weighting;  
(b) ICT: $\gamma_{d_1}=\gamma_{d_2}=10^{-5}\mathrm{eV}$, $t_{12}=10^{-5}\mathrm{eV}$, $\omega_f=5\times10^{-5}\mathrm{eV}$, $a=[10,0,10]$, $b=1$;  
(c) DRT$_1$: $\gamma_{d_1}=\gamma_{d_2}=10^{-10}\mathrm{eV}$, $t_{12}=10^{-3}\mathrm{eV}$, $\omega_f=5\times10^{-3}\mathrm{eV}$;  
(d) DRT$_2$: $\gamma_{d_1}=10^{-10}$, $\gamma_{d_2}=10^{-3}\mathrm{eV}$, $t_{12}=10^{-10}\mathrm{eV}$, $\omega_f=5\times10^{-3}\mathrm{eV}$.

\noindent
Subscripts $d_1,d_2$ denote tunnel rates to reservoirs, $12$ interdot coupling, and $\omega_f$ the resonator frequency.

\subsection*{Figure~\ref{fig:time_comparison_MSE}}

A DQD device with two gates was simulated with
\[
C_{dd} =
\begin{pmatrix}
20 & -1.5\\
-1.5 & 20
\end{pmatrix}~\mathrm{aF}, \qquad
C_{dg} =
\begin{pmatrix}
3.9 & 1.3\\
0.3 & 3.9
\end{pmatrix}~\mathrm{aF}.
\]
The tunnel coupling was fixed at $t_{12}=1\times10^{-3}\mathrm{eV}$, and harmonic confinement frequencies were set to $\omega_{x,y} = [1\times10^{-3},\,2\times10^{-3}]\mathrm{eV}$. 
Voltage-dependent tunnel couplings followed the WKB approximation with parameters $a = [10,\,10]$ and $b = 2$.

Gate voltages $V_{g1}$ and $V_{g2}$ were simultaneously swept from $0.001~\mathrm{V}$ to $0.275~\mathrm{V}$ on a $100\times100$ grid. 
All other model parameters were held constant. 
Simulations were executed using both the full Fock basis and the reduced Fock-state approach, where the maximum Fock depth was varied from 1 to 5. 

To benchmark runtime, ten randomly generated DQD devices were created with capacitance ranges 
$C_{dd} \in [10,\,30]~\mathrm{aF}$ and $C_{dg} \in [1,\,5]~\mathrm{aF}$, and tunnel couplings drawn uniformly from $[10^{-5},\,10^{-4}]\mathrm{eV}$.
Each simulation produced a $100\times100$-pixel CSD, and the mean runtime across all devices was calculated for each MFD value.

\subsection*{Figure~\ref{fig:num_dots_v_time}}

All simulations were simulated using only the CIM layer with an equal number of gates and dots  and two gates swept and
$100\times100$ gate-voltage grid with two gates swept simultaneously:
\[
V_{g1},\,V_{g2}\in[0,\,1.0]~\mathrm{V}, \quad 100~\text{points each}.
\]
The maximum dot occupancy was fixed to $N=1$ for all runs.

For each value of $M$, random capacitance matrices were generated to define the device:
$C_{dd}\in\mathbb{R}^{M\times M}$, $ C_{dg}\in\mathbb{R}^{M\times M}$.
The capacitance matrices for each DQD were generated using random symmetric configurations within realistic parameter ranges:
\[
C_{dd}^{(i,j)} \in [-5,\, -1]~\mathrm{aF}, \qquad C_{dg}^{(i,j)} \in [0.01,\,0.5]~\mathrm{aF},
\]
with diagonal elements set to 

\[
C_{dd}^{(i,i)} \in [10,\, 20]~\mathrm{aF}, \qquad 
C_{dg}^{(i,i)} \in [1,\, 5]~\mathrm{aF},
\]

For each $M\in\{2,3,\ldots,12\}$, we simulated 100 independent simulations (each with generated random matrices) and measured the time required for the simulation.

\subsection*{Figure~\ref{fig:time_comparison_2}}

Simulations were performed for both the full Fock basis and the reduced Fock-state formulations with varying maximum Fock depths (MFD = 1–3).  

Each data point corresponds to an average over one hundred randomly generated DQDs, simulated on a $100\times100$ grid in the voltage space defined by:
\[
V_{g1},\,V_{g2} \in [0,\,0.5]~\mathrm{V}.
\]
The number of gates was kept equal to the number of QDs.  
For all simulations, the tunnel coupling was fixed to $t_{12} = 1\times10^{-3}\mathrm{eV}$, the temperature was set to $T = 1.5~\mathrm{K}$, and the sensor response was computed with respect to $V_{g1}$.

The capacitance matrices for each DQD were generated using random symmetric configurations within realistic parameter ranges:
\[
C_{dd}^{(i,j)} \in [-5,\, -1]~\mathrm{aF}, \qquad C_{dg}^{(i,j)} \in [0.01,\,0.5]~\mathrm{aF},
\]
with diagonal elements set to 

\[
C_{dd}^{(i,i)} \in [10,\, 20]~\mathrm{aF}, \qquad 
C_{dg}^{(i,i)} \in [1,\, 5]~\mathrm{aF},
\]

For the full fock state simulations, the maximum charge carrier number $N$ was varied from 1 to 9.  
For the reduced fock state simulations, the Fock depth was fixed to MFD = 1, 2, and 3, and the runtime was recorded across the same set of voltage sweeps.  

\subsection*{Figure~\ref{fig:time_comparison}}

In this Figure, each layer introduces an additional level of physical realism to the underlying model of a DQD system, enabling benchmarking of the trade-offs between runtime and physical accuracy. 

All simulations were performed on 100 randomly generated DQD devices, each simulated over a $100\times100$-point voltage grid defined by
\[
V_{g1},\,V_{g2} \in [0.001,\,0.3]~\mathrm{V}.
\]
The maximum occupancy per quantum dot was fixed to one ($N=1$).

Each DQD device was generated using random symmetric capacitance matrices:
\[
C_{dd}^{(i,j)} \in [-5,\, -1]~\mathrm{aF}, \qquad 
C_{dg}^{(i,j)} \in [0.01,\, 0.5]~\mathrm{aF},
\]
with diagonal elements set to 

\[
C_{dd}^{(i,i)} \in [10,\, 20]~\mathrm{aF}, \qquad 
C_{dg}^{(i,i)} \in [1,\, 5]~\mathrm{aF},
\]
Each simulation layer was evaluated independently as follows:

\begin{itemize}
    \item \textbf{CIM:} Constant Interaction Model (no tunnelling or confinement). 
    \item \textbf{TC:} Includes interdot tunnel coupling $t_{12}\in[10^{-4},\,10^{-3}] \mathrm{eV}$. 
    \item \textbf{QC:} Adds harmonic confinement via characteristic frequencies $\omega_{x,y}\in[10^{-4},\,10^{-3}]\mathrm{eV}$. 
    \item \textbf{VDTC:} Incorporates voltage-dependent tunnel coupling using the WKB approximation with $a\in[1,\,15]$ and $b\in[1,\,5]\mathrm{eV}$.
    \item \textbf{VDC:} Enables voltage-dependent capacitances via scaling functions
    \[
    C_{dd}(V) = C_{dd}[1 + 10^{-3}\sum_i V_i]
    \]
    \[
    C_{dg}(V) = C_{dg}[1 + 5\times10^{-3}V_{g1}],
    \]
    capturing voltage sensitivity in the device electrostatics.
    \item \textbf{RF-CIM / RF-TC:} Extends the CIM and TC models with RF reflectometry readout, with dot one being used as a sensor dot and a temperature of $T = 1.0~\mathrm{K}$.
\end{itemize}

Each simulation was repeated 100 times with independently generated devices.

\subsection*{Figure~\ref{fig:SET_simulation}}
A triple-dot device with a barrier gate between dot one and two  ($P_1,P_2,P_3,B_1$) was simulated with
\[
C_{dd} =
\begin{pmatrix}
100 & -10 & -15 \\ -10 & 100 & -10 \\ -15 & -10 & 100
\end{pmatrix},\]
\[
C_{dg} =
\begin{pmatrix}
3.5 & 0.2 & 0.02 & 0.20 \\
0.2 & 3.5 & 0.20 & 0.20 \\
0.2 & 0.05 & 3.5 & 0.10
\end{pmatrix}.
\]
Static offsets: $V_{B1}=0$, $V_{P3}=0.0215$~V, with thresholds 
$V_{P1}^{\text{th}}=0.335$~V and $V_{P2}^{\text{th}}=0.35$~V.
Sweeps: $V_{P1},V_{P2}\!\in[0.10,0.55]$~V.
Noise: (i) $1/f$ with $\alpha=0.15\mathrm{W^2/Hz}$, $f_c=100 Hz$; (ii) Gaussian with $\mu=0.3\mathrm{aF}$, $\sigma=0.05\mathrm{aF}$.
Panel~(b) shows experimental reflectometry data from the same geometry.

\subsection*{Figure~\ref{fig:tunnel_rates}}
A DQD device with two gates was simulated with
\[
C_{dd} =
\begin{pmatrix}
100 & -7 \\ -7 & 100
\end{pmatrix}\mathrm{aF}, \qquad
C_{dg} =
\begin{pmatrix}
5.5 & 0.2 \\ 0.2 & 5.5
\end{pmatrix}\mathrm{aF}.
\]
Thresholds: $V_{g1}^{\text{th}}=0.01$~V, $V_{g2}^{\text{th}}=0.02$~V;
Sweeps: $V_{g1},V_{g2}\!\in[0.00,0.10]$~V.
Tunnel-rate model: $\gamma_{d_1}=5\times10^{-10}\mathrm{eV}$, $\gamma_{d_2}=1\times10^{-10}\mathrm{eV}$, 
$t_{12}=10^{-3}$, $\omega_f=5\times10^{-3}\mathrm{eV}$.
Noise: $1/f$ with $\alpha=0.01\mathrm{W^2/Hz}$, $f=1000$.
Panel~(b) shows experimental reflectometry data from the same geometry.

\subsection*{Figure \ref{fig:artefacts}}

A DQD device with two gates was simulated with
\[
C_{dd} =
\begin{pmatrix}
20 & -1.5\\
-1.5 & 20
\end{pmatrix}\mathrm{aF}, \qquad
C_{dg} =
\begin{pmatrix}
3.9 & 1.3\\
0.3 & 3.9
\end{pmatrix}\mathrm{aF}.
\]
The interdot tunnel coupling was set to $t_{12} = 1\times10^{-3} \mathrm{eV}$, and the harmonic confinement frequencies for each dot were 
$\omega_{x,y} = [1\times10^{-3},\, 2\times10^{-3}]\mathrm{eV}$.
Voltage-dependent tunnel couplings were included through the WKB approximation with coefficients $a = [10,\,10] $ and $b = 2 $.

Gate voltages $V_{g1} $ and $V_{g2} $ were swept from $0.001~\mathrm{V}$ to $0.275~\mathrm{V}$ on a $200\times200$ grid, while all other parameters were held constant. 
The simulations were performed using both the full Fock basis and the reduced Fock-state approach, with the maximum Fock depth varied from 1 to 5.

\end{document}